\documentclass[twocolumn,superscriptaddress]{revtex4-1}

\usepackage{microtype}
\usepackage{float}
\usepackage{hyperref}
\usepackage{amsmath,amsbsy, amssymb}
\usepackage{graphicx, xcolor, ulem,url}

\begin{document}
\title{Self-excited motions of volatile drops on swellable sheets}
\author{Aditi Chakrabarti}
\affiliation{John A. Paulson School of Engineering and Applied Sciences, Harvard University, Cambridge, MA 02138, USA}
\author{Gary P. T. Choi}
\affiliation{John A. Paulson School of Engineering and Applied Sciences, Harvard University, Cambridge, MA 02138, USA}
\author{L.\ Mahadevan}
\affiliation{John A. Paulson School of Engineering and Applied Sciences, Harvard University, Cambridge, MA 02138, USA}
\affiliation{Department of Physics, Harvard University, Cambridge, MA 02138, USA}
\date{\today}

\newcommand{\xx}{\mathbf{x}}

\begin{abstract}
When a volatile droplet is deposited on a floating swellable sheet, it becomes asymmetric, lobed and mobile. We describe and quantify this phenomena that involves nonequilibrium swelling, evaporation and motion, working together to realize a self-excitable spatially extended oscillator. Solvent penetration causes the film to swell locally and eventually buckle, changing its shape and the drop responds by moving. Simultaneously, solvent evaporation from the swollen film causes it to regain its shape once the droplet has moved away. The process repeats and leads to complex pulsatile spinning and/or sliding movements. We use a one-dimensional experiment to highlight the slow swelling of and evaporation from the film and the fast motion of the drop, a characteristic of excitable systems. Finally, we provide a phase diagram for droplet excitability as a function of drop size and film thickness and scaling laws for the motion of the droplet.  
\end{abstract} 

\maketitle

When a liquid drop is placed on a rigid substrate, it can either spread or round up depending on the relative magnitude of the surface energies in question. In the presence of an external gradient, the drop can move or evaporate leading to a range of dynamical phenomena that continue to enthrall and instruct, while suggesting a range of applications~\cite{Chaudhury1992,brzoska1993motions,Chaudhury2015,dos1995free,Pimienta2011,Cira2015,Liu2017,sanchez2012spontaneous}. But could a drop placed on a substrate spontaneously create and respond to gradients by itself? 

\begin{figure}[t!]
\centering
\includegraphics[width=0.42\textwidth]{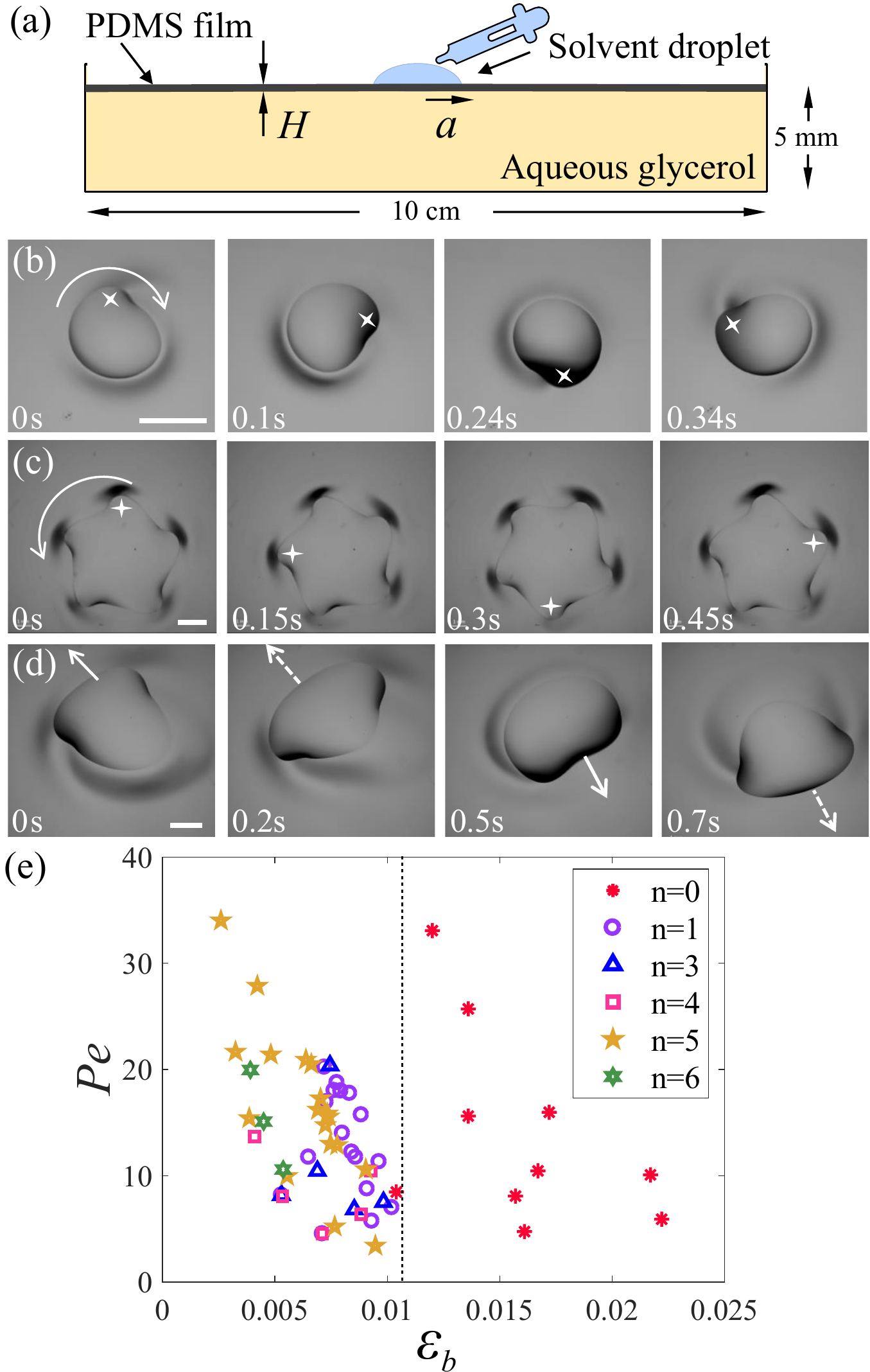}
\caption{(a) The experimental setup comprises of a thin elastic film ($H \sim$ tens of microns) made of PDMS afloat aqueous glycerol. Acetone droplets undergo spontaneous rotation as demonstrated via a single cycle of: (b) a bean mode with $n=1$ undergoing clockwise rotation (3 $\mu$l drop on 9 $\mu$m PDMS film) and (c) a star shaped mode with $n=5$ undergoing anti-clockwise rotation (20 $\mu$l drop on 19.5 $\mu$m PDMS film), and (d) a 30 $\mu$l drop of acetone undergoing oscillatory to and fro motion on a 28.5 $\mu$m PDMS film. The scale bars in (b)--(d) denote 2 mm. (e) Peclet numbers ($Pe \sim \mathcal{O}(10)$) plotted as a function of buckling strain ($\varepsilon_{b}$) for acetone droplets showing droplet instability. The dotted line indicates the critical value of $\varepsilon_{b}$, which demarcates between spherical cap states ($n=0$) and lobed states ($n=1,3,4,5,6$).}\label{figure1}
\end{figure}

Our starting point is a volatile liquid droplet of acetone (volume ranging from 1--40 $\mu$l) placed on a thin permeable membrane (thickness $H \sim$ 10--50~$\mu$m), of crosslinked polydimethylsiloxane (PDMS, Sylgard 184, 10:1) floating on aqueous glycerol (Fig.~\ref{figure1}(a)) in ambient conditions. The liquid initially forms a spherical cap on the film (Fig.~\ref{figure1}(a)), but within a few seconds, the droplet spontaneously breaks symmetry, first becoming symmetrically lobed, then chirally lobed, and finally beginning to spin (Fig.~\ref{figure1}(b) and~\cite{supplementary}, Movie~1). The number of lobes is a function of the drop size for a given film thickness; bigger drops have more lobes (Fig.~\ref{figure1}(c) and~\cite{supplementary}, Movie~1). While there is no preferred chirality, the lobes are equally probable to curve towards a clockwise or anticlockwise direction, and once it is chosen, the drop continues to spin in the same direction; however as it evaporates and becomes smaller, below a threshold size, it eventually stops. If the drop is sufficiently large, it does not spin and instead become polarized, taking the form of a kidney bean or keratocyte~\cite{bray2000cell} and can either oscillate back and forth (Fig.~\ref{figure1}(d) and~\cite{supplementary}, Movie~2) or migrate in a random direction (\cite{supplementary}, Movie~3). High speed videos of the motion of the droplets show that in some cases the drop spins smoothly, whereas in other cases it exhibits a pulsatile motion that is limited to the neighborhood of the contact line (\cite{supplementary}, Movie~4).

\begin{figure*}[t!]
\centering
\includegraphics[width=1.0\textwidth]{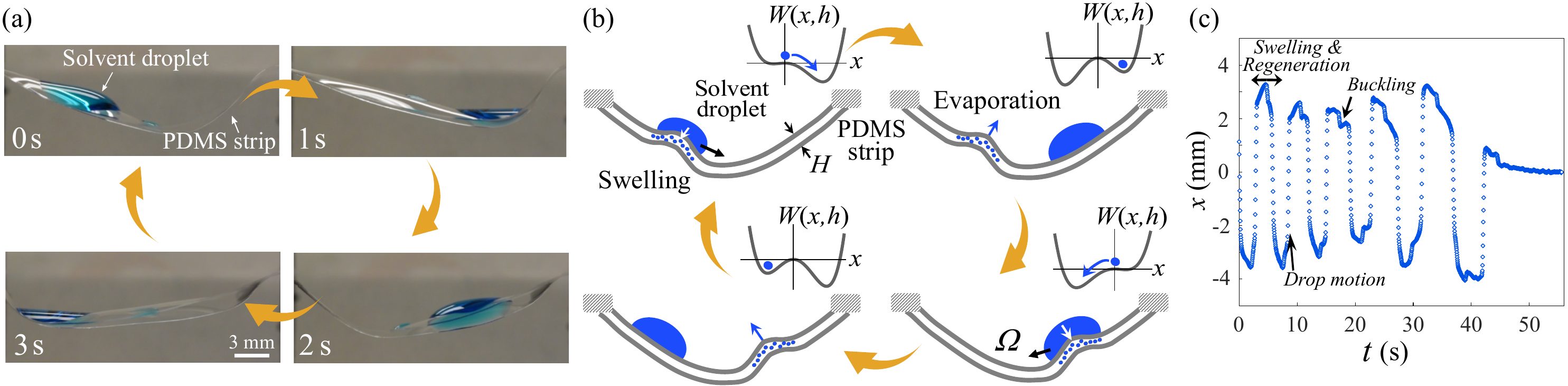}
\caption{1D excitable droplet motion. (a) A drop of volume $V=20$ $\mu$l placed on a sagging clamped PDMS strip (5 mm wide, 100 $\mu$m thick) oscillates back and forth while the film buckles periodically. (b) Schematic showing that the film locally swells in response to the solvent droplet causing the film to buckle and the droplet to slide away. The exposed swollen buckled region then loses solvent via evaporation causing the film to unbuckle. Simultaneously, the drop causes the film to swell at a new location and the process repeats. The sag in the film prevents the drop from running away and leads to an oscillating state of the drop with frequency $\Omega$. (Inset) The shape of the potential $W(x,h)$ accompanying each film state with the droplet position $x(t)$ shown with a blue dot. (c) A trace of the x-coordinate of the center of mass of an acetone droplet undergoing relaxation oscillations as a function of time, where the slow timescale is due to swelling of the PDMS film and simultaneous regeneration due to evaporation, and the fast timescale is that of the drop motion. The kinks correspond to the buckling of the film.}
\label{figure2}
\end{figure*}

Hypotheses for the undulation and motion include evaporation-driven contact line instability~\cite{wayner1993spreading} or surface tension-gradient induced Marangoni effects. Replacing acetone by other volatile liquids such as isopropanol and butanol leads to similar phenomena for the film thicknesses $H \sim $ 10--50~$\mu$m~\cite{supplementary}. However, methanol droplets deposited on the same membrane remain stationary while adopting a spherical cap shape, and hexane droplets cause the film to wrinkle but again with no accompanying motion~\cite{supplementary}. Since the droplets are composed of one solvent only, we do not expect any compositional Marangoni forces to arise. Furthermore, infrared movies (\cite{supplementary}, Movies~6--7) helped us estimate the thermal Marangoni numbers ($Ma \gg 1$, see~\cite{supplementary} for details) for the solvents used. While only acetone, isopropanol and butanol show the instability and the other liquids despite having a large thermal Marangoni number do not demonstrate the instability point to the fact that surface tension gradients are not sufficient to create the undulation and motion.

This leaves swelling-induced film deformation and the accompanying droplet motion as the most probable candidate to explain the phenomenon at hand. To quantify this, we first note that the solvent diffuses through the elastic network~\cite{FloryPaulJ1953Popc,favre1996swelling} at a rate $kE/\eta aH$, where $k$ is the permeability of drained PDMS network, $E$ is its Young's modulus, $\eta$ is the dynamic viscosity of the solvent and $a$ is the drop size. {Simultaneously, the imbibed solvent leaves the film via a rate of $J/\varepsilon_{b}H$, where $J$ is its evaporative flux and $\varepsilon_{b}\sim H/a$ is the strain associated with the fraction of the swollen film that is also responsible for film buckling.} If swelling is too fast or too slow relative to evaporation, the result would be a progression towards a (non-equilibrium) steady state. However, if the two processes are in competition, as quantified in terms of the P\'eclet number $Pe \sim \eta J a^{2}/kEH$, one can expect interesting dynamics. Estimating $Pe$ for all solvents tested for given film thicknesses show that acetone, butanol and isopropanol form lobes when $Pe \sim \mathcal{O}(10)$ whereas droplets of methanol and ethanol ($Pe < \mathcal{O}(10)$) as well as hexane and chloroform ($Pe > \mathcal{O}(10)$) do not show the instability~\cite{supplementary}. Consistent with these arguments, in environments of saturated solvent vapors, we do not observe drop oscillation or rotation, but droplet spinning resumes when vapor pressure is decreased back to ambient conditions~\cite{supplementary}.

In this sweet spot that balances evaporation, swelling and hydrodynamics, solvent imbibition in the vicinity of the drop causes the thin film ($H \sim$ tens of microns) to swell and sag to form a dimple that traps the drop (\cite{supplementary}, Movie~5), as seen in Fig.\ref{figure1}(b). If the swelling degree is too small (e.g. for a methanol droplet) no lobes form; if the swelling degree is too large (e.g. for a hexane droplet) many wrinkles form. This suggests a simple explanation for the undulation instability of the drops. An elastic strain incompatibility along the nominally circular rim of the swollen film (the region inside is swollen, while the region outside is not) causes an axisymmetric deformation mode of the film to give way to a buckled mode causing the rim to wrinkle like the edge of a leaf~\cite{liang2009shape,Holmes2011} as depicted by the light/dark regions around the lobes showing the film undulation (Fig.~\ref{figure1}); the larger the drop, the larger the perimeter so that the edge of the drop is surrounded by an undulating topography with $n$ lobes along its contact line (\cite{supplementary}, Movie~1).

 To further clarify the mechanisms at play, we turn to an even simpler realization of our observations, wherein a drop of acetone is placed on a narrow quasi-1 dimensional sagging PDMS film clamped at both ends as shown in Fig.~\ref{figure2}(a). Now the drop spontaneously oscillates back and forth while causing the film to buckle and unbuckle (\cite{supplementary}, Movie~8). To understand this, we note that the PDMS film underneath the drop swells due to solvent imbibition creating a local bulge due to buckling of amplitude $\sim \lambda$. This causes the acetone droplet to slide away from the swollen region, exposing the previously swollen region to the ambient atmosphere. The solvent then starts to evaporate from the swollen film, thereby regenerating it (Fig.~\ref{figure2}(b)). Simultaneously, the film swells at the new position of the drop causing it to buckle and pop up, causing the droplet to move back to its original position. The global sag in the PDMS strip constrains the drop to perform relaxation oscillations.
  
A minimal model for this represents the droplet as a particle at a scaled location $x(t)$ (made dimensionless using the drop size) in an asymmetric double-well potential $W(x,h)$ that oscillates slowly so that its minimum switches from one side to another, causing the particle to follow the minimum. The simplest form of this potential is $W(x,h) = \mu (V(x) -h(t) x),$ where $V(x)=x^4/12 - \beta x^2/2$ is associated with the double-well, and the last bilinear term with a dynamically varying tilt $h(t)$ due to the simultaneous effect of (i) the location of the drop which causes the film to swell locally, and (ii) evaporation from the previous location of the drop, as shown in Fig.~\ref{figure2}(b). There are three time scales in the problem, that associated with gravitational motion of the drop $\tau_x \sim \eta_{film}/ \rho g a$, that due to swelling-induced buckling with time $\tau_s \sim  \eta H^2/kE$, and evaporation $\tau_e \sim \varepsilon_{b}H/J$. In the simplest setting, we choose the larger of the latter two with $\tau_h ={\rm max} [\tau_s, \tau_e]$ as it defines the rate of tilting the potential, and define a ratio $\mu = \tau_h/\tau_x $. Then, in the limit of overdamped dynamics, the particle (drop) moves according to $ \dot x = \mu (h - F(x))$, where $F(x) = V'(x)= x^3/3- \beta x$. Since the tilt itself evolves slowly due to evaporation from the exposed film and swelling of the film due to the droplet, we approximate this via the simple linear dynamical law $ \mu \dot h = -x $. When $\mu \ne 0, \beta>0$, the pair of equations is just the canonical Van der Pol equation for self-excited dynamics~\cite{vanderpol1926}, $  \ddot x- \mu \dot x (\beta - x^2)+  x = 0$. For the weakly non-linear case, when $\mu  \ll 1$, we get almost periodic oscillations~\cite{supplementary}, while in the limit $\mu  \gg 1$ (Fig.~\ref{figure2}(c)), we get relaxation oscillations with slow-fast dynamics; then the drop moves quickly, with $\dot x \sim \mu$, but the overall tilt changes slowly with a speed $1/\mu$.  
  
In dimensional terms, this implies that the period of the oscillations is determined by the larger of two time scales: that required to swell the film, i.e. $\tau_s \sim  \eta H^2/kE$ ($\sim 1s$), until it buckles with a critical strain driven by the balance between bending and stretching~\cite{landau1959theory}, i.e. $\varepsilon_{b}\sim H/a$, and that associated with the evaporation of the drop. When the drop slides away, evaporation from the swollen bulge occurs over a timescale of $\tau_{e}\sim \varepsilon_{b}H/J$ ($\sim 5s$). Since the swelling and evaporation driven regeneration occur simultaneously, the slow timescale is of order $\tau_{e}$. Over time, the drop itself shrinks due to evaporation, and the time period of oscillations increases and finally the drop stops moving when it reaches a critical size where the swelling occurs on a small enough scale that is insufficient to buckle the film (\cite{supplementary}, Movie~8).

\begin{figure*}[t!]
\centering
\includegraphics[width=0.9\textwidth]{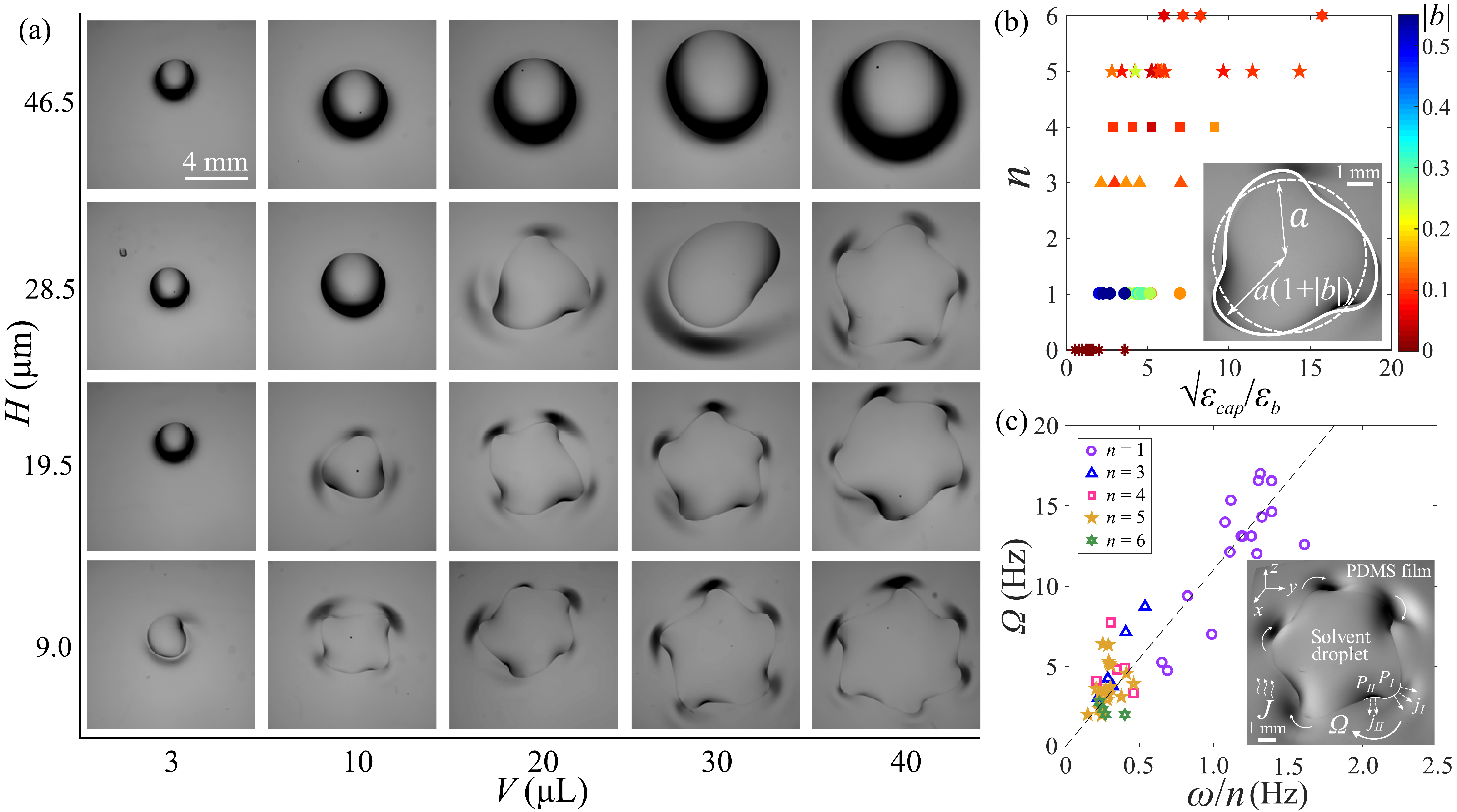}
\caption{(a) Phase space of instantaneous snapshots of droplets of acetone (volume $V$) when they are deposited on PDMS films (thickness $H$). Smaller droplets on thicker films are ``pinned'' in spherical cap state whereas larger droplets on thinner films spontaneously form lobes $n$ ($=1$ to 6) that undergo rotation. (b) Bifurcation diagram of the shapes comprising spherical pinned state ($*$) with $n=0$ and chiral spinning state depicting number of lobes $n$ as a function of the dimensionless elastocapillary number ($\sqrt{\varepsilon_{cap}}/\varepsilon_{b}$). The color of each filled symbol depicts the amplitude of the lobes normalized by the radius of the drops ($|b|$). There is a critical threshold around $\sqrt{\varepsilon_{cap}}/\varepsilon_{b} \simeq 3$ where the shapes bifurcate from spherical cap to lobed state. (Inset) An example of a fitted shape ($n=3$) with contact radius $a$ and amplitude $|b|$ labelled. {(c) Frequency ($\Omega$) of the spinning drops in the 3D case scales as $\omega/n$, where $\omega \sim 1/\sqrt{\tau_{x} \tau_{s}}$, where the black dotted line is a least squares fit to the data with slope 11. (Inset). Schematic of the solvent droplet ($n=5$) that swells the PDMS film leading to undulations, which are enhanced here for visualization, and $J$ is the evaporative flux from the film. The liquid in the lobes traverse from hills to valleys along the circumference, giving rise to coordinated spinning of the drop with frequency $\Omega$.}}
\label{figure3}
\end{figure*}

Having understood the basic mechanisms in the simple 1D system, we now quantify the phase space of the droplet shapes, determined by the droplet volume and the PDMS film thickness on the 2D films. On film thicknesses $H <20$ $\mu$m, acetone droplets ($V \sim 3$--$40$~$\mu$L) form $n$-lobed chiral shapes that spin spontaneously (Fig.~\ref{figure3}(a)). On thicker films ($H >40$ $\mu$m), small and large drops form spherical caps that stay pinned. The tendency for the droplet to break symmetry and move, can be parameterized by the ratio of drop size $a$ to the elastocapillary length $\ell= \sqrt{EH^3/T}$, which characterizes the substrate softness (in terms of the tension $T\sim EH\varepsilon_{cap}$, where $\varepsilon_{cap}$ is the elastic strain in the film due to wetting of the drop). This is denoted by the elastocapillary number $a/\ell \sim \sqrt{\varepsilon_{cap}}/ \varepsilon_{b}$, where $ \varepsilon_{b}$ is the buckling strain. The shape of the lobed drops can be quantified using a simple polar representation~\cite{Sabrina2018} with coordinates $(r(s), \theta(s))$ in the domain $s \in [-\pi, \pi]$, where $r(s) = a(1+b \cos(n s))$ and $\theta(s) = s + \frac{c}{n} \cos(ns + \psi) + \phi$. Here, $a$ represents the overall radius of the drop, $b$ represents the non-dimensional amplitude of the lobes, $c$ defines the asymmetry of the lobes, $\psi$ accounts for the local phase shift at the lobes and $\phi$ accounts for the global phase shift (Fig.~\ref{figure3}(b) inset;~\cite{supplementary}). The shape of the droplets are similar to those of cell fragments and primitive cells~\cite{bray2000cell} - hardly surprising as the first few unstable modes of active drops always take the same geometric forms.

Plotting the number of lobes ($n$) of the experimentally observed shapes as a function of the elastocapillary number reveals that when $\sqrt{\varepsilon_{cap}}/ \varepsilon_{b} \ge 3$, i.e. when the drop is ``soft'' enough, its contact line becomes asymmetric and forms lobes (Fig.~\ref{figure3}(b)). As the drop settles on the soft sheet with an undulating rim, the film swells and buckles to form saddle-like structures that then become asymmetric (and chiral) around the rim (Fig.~\ref{figure3}(a)), and start to spin in a coordinated way. When $n<6$, they synchronize and rotate, while when $n>7$, the lobes fail to synchronize, and quiver instead (\cite{supplementary}, Movie~9). Introducing neutrally buoyant hollow glass spheres in the droplet shows that the particles only move in a boundary layer near the contact line when $n \in [2,6]$ (\cite{supplementary}, Movie~10), whereas for drops with $n \sim 1$, particles move with the liquid showing global rotation of the drop. Near the contact line, variations in solvent vapor concentration from both the wetted film and the drop can lead to gradients that can cause the lobes of the droplet to break symmetry and become chiral. In all the experiments, the chirality and direction of motion of each lobe are correlated: the liquid moves from the convex side of the lobe towards its concave side (\cite{supplementary}, Movie~1,4). To understand why this is so, we recall the Kelvin relation~\cite{thomson18724} that suggests it is easier to evaporate from convex surfaces than from concave surfaces as the vapor pressure due to positive curvature ($P_{I}$) is higher than that due to negative curvature ($P_{II}$). For the chiral lobes observed in our system, this would imply asymmetric evaporation from the convex and concave sides of the lobe, leading to motion from the high evaporation (convex) to the low evaporation (concave) ($j_{I} > j_{II}$), provided the contact line is mobile~\cite{man2017vapor} (Fig.~\ref{figure3}(c) Inset). Since the contact line of the droplet in our system sits on a saturated film, it is minimally pinned. Together, the asymmetric evaporation and the synchronized motion of lobes from convex to concave sides of the lobes give rise to an overall rotation of the drops on the PDMS film.

To understand how to generalize our understanding of the 1D oscillations of the drop to the 2D spinning drop, we note that along the rim of the drop strong gradients in swelling cause sheet to form localized wrinkles that create a periodic undulatory landscape near the rim. Spatial variations in liquid evaporation from the lobes as well as the wetted film around the drop can spontaneously break chiral symmetry of the lobes causing the rim of the drop to move tangentially - which then drives the wrinkles to rearrange via evaporation and swelling. If the number of lobes is small, the dynamics of the liquid in the vicinity of the contact line is analogous to the 1D system, except that it is periodic. If the number of lobes is large, it becomes hard to coordinate the motion of the droplet edge, leading to a shivering frustrated drop (\cite{supplementary}, Movie~9). 

Quantitatively, the frequency of a complete rotation in the periodic case for small values of $\mu$, $\Omega \sim 1/n\sqrt{\tau_{s}\tau_{x}}$, where the liquid moves from one lobe to the next. We find that our experiments are consistent with this simple scaling law, as shown in Fig.~\ref{figure3}(c) where the angular frequency of the droplets with varying lobe numbers collapses for the relationship $\Omega \sim \omega/n $, where $\omega \sim 1/\sqrt{\tau_{s}\tau_{x}}$ is the characteristic frequency. In any single experiment, the drop continues to spin until its size reduces below a critical threshold determined by the parameter $\sqrt{\varepsilon_{cap}}/ \varepsilon_{b} \sim 1.8$ in Movie~2~\cite{supplementary}, a value close to the threshold observed in Fig.~\ref{figure3}(b).

Our experiments have shown that a volatile drop on a soft responsive substrate can create and respond to local deformation and evaporation gradients and lead to spontaneous oscillations. Three non-equilibrium processes: evaporation, solvent flow and solvent-driven swelling of a thin elastic film conspire to produce the oscillations, and one might have thought that the parameter space where these work together is small. We have shown that this is not the case - a range of solvent types and droplet sizes on thin responsive substrates satisfy the conditions for the phenomena to be observable, i.e. that $Pe \sim \mathcal{O}(10)$, and the substrate be easy to buckle (i.e. thin enough). Together these processes drive self-excited motion in drops over a robust range of parameters. Harnessing such instabilities and motion in thin film systems might provide a natural way to drive small scale engines building on recent work in this domain~\cite{ma2013bio,chung2014evaporative,chen2015scaling}.

{\bf Acknowledgment} We thank Manoj K. Chaudhury for fruitful discussions. This work was supported in part by the Croucher Foundation, the Harvard Quantitative Biology Initiative and the NSF-Simons Center for Mathematical and Statistical Analysis of Biology at Harvard, Grant No. 1764269 (to G. P. T. C.), and NSF DMR 14-20570 MRSEC, NSF DMR 15-33985 Biomatter and NSF CMMI 15-36616 (to L. M.).


%\bibliographystyle{apsrev4-1}
%\bibliography{spinningbib}

\end{document}